\documentclass[twocolumn,secnumarabic,amssymb, 
amsmath,nobibnotes, aps, showpacs,prl]{revtex4}

\usepackage{graphicx}
\usepackage{bm}

\begin{document}
\title{CMB Polarization Systematics Due to Beam Asymmetry:\\
Impact on Inflationary Science}
\ \\
\author{Meir Shimon $^{1}$, Brian Keating$^{1}$, Nicolas Ponthieu$^{2}$, Eric Hivon$^{3,4}$}
\affiliation{
{${}^{1}$}Center for Astrophysics and Space Sciences, University
of California, San Diego, 9500 Gilman Drive, La Jolla, CA, 92093-0424\\
{${}^{2}$}Institut d'Astrophysique Spatiale, Bat. 121, Universite Paris XI, 91405
Orsay Cedex, France\\
{${}^{3}$}California Institute of Technology, Pasadena CA 91125, U.S.A.\\
{${}^{4}$}Institut d'Astrophysique de Paris, 98 bis boulevard Arago, F-75014 Paris, France}
\date{April 7, 2008}


\begin{abstract}
Cosmic microwave background (CMB) polarization provides a unique window into cosmological inflation; 
the amplitude of the B-mode polarization from last scattering is
uniquely sensitive to the energetics of inflation. However, numerous
systematic effects arising from optical imperfections can contaminate the
observed B-mode power spectrum. In particular, 
systematic effects due to the coupling of the underlying temperature and polarization
fields with elliptical or otherwise asymmetric beams yield
spurious systematic signals. This paper presents a 
non-perturbative analytic calculation of some of these signals. 
We show that results previously derived in real space can be generalized, 
formally, by including infinitely many higher-order corrections to the 
leading order effects. These corrections 
can be summed and represented as analytic functions 
when a fully Fourier-space approach is adopted from the outset. 
The formalism and results presented in this paper were created to 
determine the susceptibility of CMB polarization probes of the 
primary gravitational wave signal but can be easily extended to the 
analysis of gravitational lensing of the CMB. 
\end{abstract}

\maketitle

\section{Introduction}
Upcoming cosmic microwave background (CMB) polarization experiments are poised to immensely 
improve our understanding of the early universe. 
The significance of polarization lies not only in the fact 
that it increases the amount of data provided by temperature anisotropy
alone but also because it is more sensitive to several physical processes which took 
place in the early universe, e.g. the gravitational wave background
produced during inflation (Seljak \& Zaldarriaga [1], 
Kamionkowski, Kosowsky \& Stebbins [2]) and reionization 
(Zaldarriaga [3], and Fan, Carilli \& Keating [4]). 
Partially (but not only) due to this 
fact, CMB polarization has a unique dependence on the basic 
cosmological parameters. This feature can be used to remove 
some of the degeneracies afflicting cosmological parameter estimation 
from temperature anisotropy alone. 
One of the CMB polarization's main features is the 
dependence of its B-mode (curl-like polarization) 
on the amplitude of the stochastic gravitational wave background generically 
predicted by inflationary models. 
Due to its faintness, the B-mode polarization is 
prone to degradation by various systematic effects on a wide range of 
scales and it is important to remove these spurious contributions. 
This must be done {\it in addition} to controlling the various 
systematics induced by diffuse galactic 
emission (Amblard, Cooray \& Kaplinghat [5]) 
and contamination from E-B mixing due to partial sky coverage 
(Lewis, Challinor \& Turok [6], de Oliveira-Costa \& Tegmark [7], 
Brown, Castro \& Taylor [8]). 

This paper describes an analytic approach to
assess systematics induced by imperfections 
of the polarimeter's main beam.  
Several effects which contaminate the power spectra 
stem from temperature and polarization variations 
over scales comparable to the beamwidth 
(Hu, Hedman \& Zaldarriaga [9], Ponthieu [10], Rosset et al. [11], 
O'Dea, Challinor \& Johnson [12]); 
these effects can be modelled and characterized by the spurious 
$C_{l}$ they produce.

Our results are power spectra presented as a combination of the underlying 
power spectra with mixing coefficients which are infinite sums of analytic 
functions.
In practice however, this infinite series must be truncated and therefore 
our result is effectively equivalent to a series expansion.
As expected, for small beam imperfections we find that 
the higher order corrections contribute negligibly to the spurious polarization.
We define the small parameters characterizing the systematic effects 
in Table I. They are; the gain factor $g$, the differential
beamwidth of the beams $\mu$, the differential pointing $\rho$, 
the beam ellipticity $e$ and the beam rotation $\varepsilon$. 

This paper is organized as follows; in section 2 we present 
the basic mathematical formalism of spin-weighted fields used to 
characterize the systematic effects, beam
convolution, and our analytic results for the temperature-polarization leakage
and polarization conversion in a single beam. 
Ultimately, we consider bolometric polarimetry which is most 
conveniently described using the Stokes parameters (e.g, Masi et al. [13])
$I$, $Q$ and $U$ 
($V=0$ for CMB polarization) as opposed to the Jones matrix formalism which 
is particularly useful for describing coherent polarimeters (Hu, Hedman \& 
Zaldarriaga [9], O'Dea, Challinor \& Johnson [12]). 
These parameters are derived from differences in intensity in the 
Gaussian 2-D polarized beam response function for each polarization.
Our two-beam
experiment and the induced power spectra are derived in section 3 and the
numerical results for the B-mode power
spectrum are described in section 4. 
We end with a discussion of the impact of these effects in Section 5.

\section{Mathematical Formalism}

We work entirely in Fourier space and begin with the 
expansion of both temperature and $Q$ and $U$ Stokes parameters in 
plane waves. Since all the effects considered here are due to the beam 
asymmetry, and the beamwidths
are typically on the degree or sub-degree scale, we can safely employ the 
flat-sky approximation as far as effects related to 
the beam shape and size are considered. 
The underlying physical power spectrum is 
calculated by CAMB (Lewis, Challinor \& Lasenby [14]) using the full sky. 
The temperature and other (integer-spin) combinations 
of the Stokes parameters are expanded in harmonic space as 
follows (e.g. Zaldarriaga \& Seljak [15])
\begin{eqnarray}
T(\hat{n})&=&\sum_{l,m}a_{lm}Y_{lm}(\hat{n})\nonumber\\
(Q\pm iU)(\hat{n})&=&\sum_{l,m}a_{\pm 2,lm}\ _{\pm 2}
Y_{lm}(\hat{n})
\end{eqnarray}
where the expansion coefficients of the spin $\pm 2$ polarization parameters
can be presented in terms of the $E$ and $B$ polarization modes
\begin{eqnarray}
a_{\pm 2,lm}=E_{lm}\pm iB_{lm}.
\end{eqnarray}
$E$ and $B$ are scalar and pseudoscalar under parity (having even and odd
parities) respectively, and are sometimes referred to as the `electric' and
`magnetic' (or `gradient' and `curl') polarization components.
In the flat sky approximation, Eq.(1) becomes
\begin{eqnarray}
T({\bf x})&=&\frac{1}{2\pi}\int T_{{\bf l}}e^{i{\bf l}\cdot{\bf x}}d^{2}{\bf
l}\nonumber\\
X^{\pm}({\bf x})&\equiv &(Q\pm iU)({\bf x})\nonumber\\
&=&\frac{1}{2\pi}\int(E_{{\bf l}}\pm iB_{{\bf
l}})e^{i{\bf
l}\cdot{\bf x}}e^{\pm 2i(\phi_{{\bf l}}-\phi_{{\bf x}})}d^{2}{\bf l}
\end{eqnarray}
where $T_{{\bf l}}$, $E_{{\bf l}}$ and $B_{{\bf l}}$ are the Fourier 
components which are functions of the wave vector ${\bf l}$ only. 

Since in real space the temperature and polarization patterns are convolved
with the beams,
these expressions are simply the product of their Fourier transforms in 
Fourier space. We restrict the discussion to an elliptical gaussian beam 
(with major and minor axes $\sigma_{x}$ and $\sigma_{y}$)
\begin{eqnarray}
B({\bf x})=\frac{1}{2\pi\sigma_{x}\sigma_{y}}\exp\left(-\frac{(x-\rho_{x})^{2}}{2\sigma_{x}^{2}}
-\frac{(y-\rho_{y})^{2}}{2\sigma_{y}^{2}}\right)
\end{eqnarray}
and its Fourier transform is
\begin{eqnarray}
\tilde{B}({\bf l})=
\exp\left(-\frac{l_{x}^{2}\sigma_{x}^{2}}{2}-\frac{l_{y}^{2}\sigma_{y}^{2}}{2}
+i{\bf l}\cdot{\bf\rho}\right).
\end{eqnarray}
The pointing error merely shifts the phase of the beam representation 
in Fourier-space.
It is useful to switch to polar coordinates at this point
\begin{eqnarray}
l_{x}&=&l\cos(\phi_{{\bf l}}+\psi-\alpha)\nonumber\\
l_{y}&=&l\sin(\phi_{{\bf l}}+\psi-\alpha)\nonumber\\
\rho_{x}&=&\rho\cos\theta\nonumber\\
\rho_{y}&=&\rho\sin\theta
\end{eqnarray}
where the angles $\psi$, $\alpha$ and $\theta$ are defined below.
The Fourier representation of the beam (Eq.5) then becomes
\begin{eqnarray}
\tilde{B}({\bf l})d^{2}{\bf l}=e^{-y-z\cos 2(\phi_{l}+\psi-\alpha)
+il\rho\cos(\phi_{l}-\alpha-\theta+\psi)}ldld\phi_{l}
\end{eqnarray}
where
\begin{eqnarray}
y&\equiv&\frac{l^{2}}{4}(\sigma_{x}^{2}+\sigma_{y}^{2})\nonumber\\
z&\equiv&\frac{l^{2}}{4}(\sigma_{x}^{2}-\sigma_{y}^{2}).
\end{eqnarray}
The definitions of the parameters in terms of the mean beamwidth $\sigma$, differential 
beamwidth $\mu$, and ellipticity $e$, are given in Table II.
Employing the expansion of 2-D plane waves in terms of cylindrical Bessel
functions
\begin{eqnarray}
e^{il\rho\cos (\phi_{l}-\phi_{\rho})}=\sum_{n=-\infty}^{n=\infty}
i^{n}J_{n}(l\rho)e^{in(\phi_{l}-\phi_{\rho})},
\end{eqnarray}
the definition of modified Bessel function
\begin{eqnarray}
I_{n}(z)=i^{-n}J_{n}(iz),
\end{eqnarray}
and the symmetry relation
\begin{eqnarray}
J_{-n}(z)=(-1)^{n}J_{n}(z),
\end{eqnarray}
Eq.(7) becomes
\begin{eqnarray}
\tilde{B}({\bf l})&=& e^{-y}
\sum_{n=-\infty}^{\infty}\sum_{m=-\infty}^{\infty}i^{2m+n}I_{m}(z)
J_{n}(l\rho)\nonumber\\
&\times &e^{i(2m+n)\psi-in\theta}e^{i(2m+n)(\phi_{l}-\alpha)}\nonumber\\
&\equiv &\sum_{n=-\infty}^{\infty}\sum_{m=-\infty}^{\infty}B_{m,n}e^{i(2m+n)(\phi_{l}-\alpha)}
\end{eqnarray}
where $\alpha\equiv\beta+\theta+\psi$ is the angle 
of the polarization axis in some coordinate system 
fixed to the sky (Fig. 1). 
We will employ this relation repeatedly in this work.

\begin{figure*}[h!t!p!]
\includegraphics[width=10.5cm]{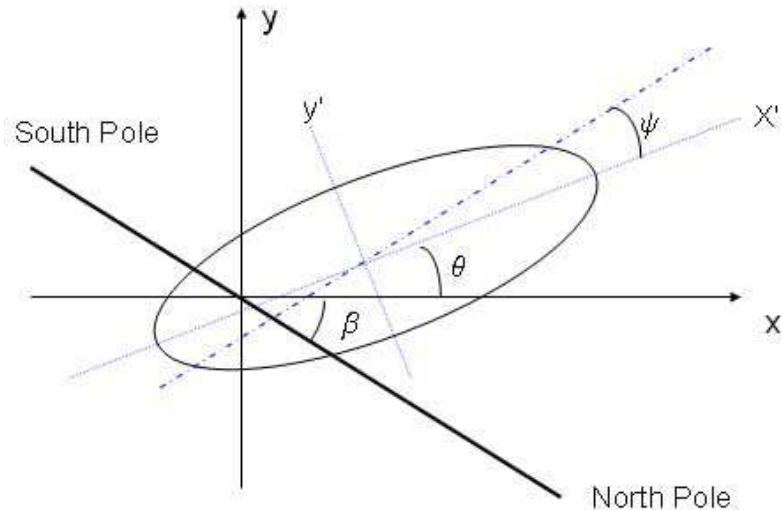}
\caption{The beam profile with the angles $\beta$, $\theta$ and $\psi$ shown for a single beam.
The horizontal x and vertical y axes are fixed to the focal plane. The axis of polarization sensitivity 
makes an angle $\psi$ with the ellipse major axis.}
\end{figure*}

\begingroup
\begin{table}[h!t!]
\begin{tabular}[c]{|c|c|c|}
\hline
~effect~&~parameter~&~definition~\\
\hline
~gain~& $g$ & $g_{1}-g_{2}$ \\
\hline
~monopole~& $\mu$ & $\frac{\sigma_{1}-\sigma_{2}}{\sigma_{1}+\sigma_{2}}$ \\
\hline
~dipole~& $\rho$ & ${\bf \rho}_{1}-{\bf \rho}_{2}$ \\
\hline
~quadrupole~& $e$ & $\frac{\sigma_{x}-\sigma_{y}}{\sigma_{x}+\sigma_{y}}$ \\
\hline
~rotation~& $\varepsilon $ & $\frac{1}{2}(\varepsilon_{1}+\varepsilon_{2})$ \\
\hline
\end{tabular}
\caption{Definitions of the parameters associated with the systematic effects.
Subscripts 1 and 2 refer to the first and second beams, respectively. 
The differential gain parameter $g$ and the beam rotation parameter $\varepsilon$ 
are not related to the beam shape and therefore are not defined in Eqs. (7)-(8), 
but rather are global parameters which define beam mismatch. 
We defer the exact definitions of these parameters to the relevant sections.} 
\end{table}
\endgroup

\section{Dual Polarized Beam Experiment}

We represent the data collected by the detectors as (e.g. Tegmark [16])
\begin{eqnarray}
d({\bf p})=A({\bf p})m({\bf p})+n({\bf p})
\end{eqnarray}
where $d$ is the measured data, $A({\bf p})$, $m({\bf p})$ and $n({\bf p})$ 
are, respectively, the pointing matrix, map vector with coefficients $T$, $Q-iU$, $Q+iU$, 
and the noise, at the pixel ${\bf p}$. 
For gaussian white noise, the optimal map $\tilde{m}$ assumes the form (Tegmark [16])
\begin{eqnarray}
\tilde{m}({\bf p})=\left(\sum_{j\in {\bf p}} A_{j}^{T}A_{j}\right)^{-1}
 \left(\sum_{j\in {\bf p}}A_{j}^{T}d_{j}\right)
\end{eqnarray}
where the sums run over all measurements of the pixel ${\bf p}$, $A$ is given by
\begin{eqnarray}
A_{j}=(1,\frac{1}{2}e^{2i\alpha_{j}},\frac{1}{2}e^{-2i\alpha_{j}}),
\end{eqnarray}
and $A^{T}$ is $A$ transposed.
In general the matrix elements of $A$ depend on the angle
$\alpha\equiv \beta+\theta+\psi+\epsilon$ where the angles $\theta$ and $\psi$ 
were defined above, the angle $\epsilon$ is the sum of uncertainties in these 
two angles, and $\beta$ is the angle between some arbitrary axis and 
the $x$ axis of the focal plane.
Clearly, the angular coverage uniformity of each pixel 
depends on the details of the scanning strategy.
We address this issue in the following section.

\subsection{General Considerations}

The effects we consider in this work arise either from 
circular beams with unmatched main-beam full width at half maximum (FWHM; 
called the {\it monopole} effect) or from beams with differential ellipticities 
({\it quadrupole} effect).
The effect of beam ellipticity on the temperature anisotropy power spectrum 
was considered earlier by, e.g. Souradeep et al. [17], 
and the effect of beam asymmetry on the two-point correlation functions 
of the temperature anisotropy and polarization was considered by Ng [18].
Our present work, however, is multipole space based. 
Also, differential {\it gain} 
or differential pointing ({\it dipole} effect) can induce further spurious
polarization signals from temperature leakage due to beam mismatch 
as will be discussed below. Finally, if the polarization sensitivity 
axes of the beams are rotated 
(differential {\it rotation} effect) we expect mixing 
between the polarization $E$ and $B$ modes, and associated 
leakage of power between $E$ and $B$. 

We first consider a general case with $\varepsilon=0$ (i.e., no rotation error). 
In this case, Eq.(14) for the two beam experiment reads
\begin{eqnarray}
{\bf I'}({\bf p})
&=&[(A^{T}A)+R_{\pi}(A^{T}A)R^{-1}_{\pi}]^{-1}_{{\bf p}}\nonumber\\
&\times &\left[(A^{T}A)
{\bf I}_{1}({\bf p})
+R_{\pi}(A^{T}A)R^{-1}_{\pi}
{\bf I}_{2}({\bf p})\right]
\end{eqnarray}
where the vector ${\bf I}({\bf p})\equiv[T,Q-iU,Q+iU]({\bf p})$,
the subscripts $1$ and $2$ refer to the first and second beams, respectively, 
and $R_{\pi}$ is the rotation matrix by $\pi$.
Using Eq.(15), we obtain
\begin{eqnarray}
T'&=&\langle T_{+}\rangle+\frac{1}{2}\langle(Q_{-}-iU_{-})e^{2i\alpha}\rangle\nonumber\\
&+&\frac{1}{2}\langle(Q_{-}+iU_{-})e^{-2i\alpha}\rangle\\
Q'\pm iU'&=&\frac{1}{D}\langle Q_{+}\pm iU_{+}\rangle+\frac{2}{D}\langle T_{-}e^{\pm 2i\alpha}\rangle\nonumber\\
&+&\frac{1}{D}\langle(Q_{+}\mp U_{+})e^{\pm 4i\alpha}\rangle\nonumber\\
&-&\frac{1}{D}\langle Q_{+}\mp iU_{+}\rangle\langle e^{\pm 4i\alpha}\rangle\nonumber\\
&-&\frac{2}{D}\langle T_{-}e^{\mp 2i\alpha}\rangle\langle e^{\pm 4i\alpha}\rangle\nonumber\\
&-&\frac{1}{D}\langle(Q_{+}\pm iU_{+})e^{\mp 4i\alpha}\rangle\langle e^{\pm 4i\alpha}\rangle
\end{eqnarray}
where 
\begin{eqnarray}
X_{\pm}\equiv\frac{1}{2}(X_{1}\pm X_{2})
\end{eqnarray}
with $X=T,Q$ or $U$ and $D\equiv 1-\langle e^{4i\alpha}\rangle \langle e^{-4i\alpha}\rangle$.
Angular brackets stand for averaging the trigonometric functions due to the scanning strategy.
To simplify the following discussion we encapsulate the properties of the scanning strategy 
\begin{eqnarray}
f(m,n)&&\equiv\langle e^{-i(2m+n)\alpha}\rangle\nonumber\\
h_{\pm}(m,n)&&\equiv\frac{1}{D}[f(m,n)-f(m\pm 2,n)\langle e^{\pm 4i\alpha}\rangle].
\end{eqnarray}
As an example of $f$ and $h_{\pm}$, for an ideal scan strategy, 
the angle $\alpha$ is uniformly sampled, and as can be 
seen from Eq.(20), only combinations of $n$ and $m$ which satisfy $2m+n=0$ 
result in non-vanishing $f(m,n)$ and $h_{\pm}(m,n)$. When the scanning strategy is non-ideal, and 
the beam mismatch does not have the required quadrupole symmetry, the above functions 
$f(m,n)$ and $h_{\pm}(m,n)$ couple to the beam mismatch to satisfy the necessary 
properties of spin $\pm 2$ fields 
for certain combinations of $m$ and $n$ (see Eq. 27 below). 
Employing Eq.(3) we obtain the temperature anisotropy and polarization modes in Fourier space
\begin{eqnarray}
&&\tilde{T}'_{{\bf l}}=
\sum_{m,n}[(B_{+})_{m,n}e^{i(2m+n)\phi_{l}}\tilde{T}_{{\bf l}}]\star\tilde{f}_{{\bf l}}(m,n)\nonumber\\
&&+\frac{1}{2}\sum_{m,n}[(B_{-})_{m+1,n}e^{i(2m+n)\phi_{l}}(\tilde{E}_{{\bf l}}-i\tilde{B}_{{\bf l}})]\star\tilde{f}_{{\bf l}}(m,n)\nonumber\\
&&+\frac{1}{2}\sum_{m,n}[(B_{-})_{m-1,n}e^{i(2m+n)\phi_{l}}(\tilde{E}_{{\bf l}}+i\tilde{B}_{{\bf l}})]\star\tilde{f}_{{\bf l}}(m,n)\nonumber\\
&&\tilde{E}'_{{\bf l}}\pm i\tilde{B}'_{{\bf l}}=
\sum_{m,n}[(B_{+})_{m,n}e^{i(2m+n)\phi_{l}}(\tilde{E}_{{\bf l}}\pm i\tilde{B}_{{\bf l}})]\star\tilde{h_{\pm}}_{{\bf l}}(m,n)\nonumber\\
&&+\sum_{m,n}[(B_{+})_{m\pm 2,n}e^{i(2m+n)\phi_{l}}(\tilde{E}_{{\bf l}}\mp i\tilde{B}_{{\bf l}})]\star\tilde{h_{\pm}}_{{\bf l}}(m,n)\nonumber\\
&&+2\sum_{m,n}[(B_{-})_{m\pm 1,n}e^{i(2m+n)\phi_{l}}\tilde{T}_{{\bf l}}]\star\tilde{h_{\pm}}_{{\bf l}}(m,n)
\end{eqnarray}
where $\star$ stands for convolution in Fourier space 
and $(B)_{m,n}$ are defined in Eq.(12).

The systematic power spectra are obtained by taking the
angular average (in Fourier space) of the squared 
modulus of the expressions in Eq.
(21). Due to the assumed statistical isotropy of the underlying power
spectra, the celestial temperature and polarization can be taken out of the
integral
\begin{eqnarray}
C'^{XY}_{l}\equiv\langle XY^{*}\rangle=\int 
XY^{*}\frac{d\phi_{l}}{2\pi}
\end{eqnarray}
and we are left with integrations over the beam profiles in Fourier space.
We use the auxiliary quantities
\begin{eqnarray}
A&&\equiv\langle T_{{\bf l}}(E-iB)^{*}\rangle\nonumber\\
C_{\pm}&&\equiv\langle(E+iB)(E\pm iB)^{*}\rangle
\end{eqnarray}
in terms of which the power spectra are
\begin{eqnarray}
C'^{E}_{l}&=&\frac{1}{2}{\rm Re}(C_{+}+C_{-})\nonumber\\
C'^{TE}_{l}&=&{\rm Re}(A)\nonumber\\
C'^{B}_{l}&=&\frac{1}{2}{\rm Re}(C_{+}-C_{-})\nonumber\\
C'^{TB}_{l}&=&{\rm Im}(A)\nonumber\\
C'^{EB}_{l}&=&\frac{1}{2}{\rm Im}(C_{-})
\end{eqnarray}
and
\begin{eqnarray}
C'^{T}_{l}=\langle|\tilde{T'}_{{\bf l}}|^{2}\rangle.
\end{eqnarray}
The explicit forms of $A$ and $C_{\pm}$ are given in the Appendix.

Two special cases, for which these results simplify considerably, 
are an ideal isotropic scan (in which every pixel is being scanned 
many times in a random orientation so both $f(m,n)$ and $h_{\pm}(m,n)$ 
identically vanish (except for the case $2m+n=0$) as do their Fourier transforms 
$\tilde{f}({\bf l})$ and $\tilde{h}_{\pm}({\bf l})$), 
and uniform coverage of the observed field 
(the number of hits per pixel is not large so $f$ and $h_{\pm}$ get constant, non-zero 
values). In this case the Fourier transforms of $f$ and $h_{\pm}$ are $\delta$-functions 
and there is no multipole-mixing 
in Eq.(21). Clearly, the first case is the limit of the second when $N_{{\rm hits}}$ is very large.
In principle, the field of view can be few angular degrees so one may wonder if the convolution 
in Eq.(21) in Fourier rather than harmonic space is warranted. Indeed, all the underlying power spectra 
peak at high multipoles and the effects we consider here are on scales of few percent of the beamwidth which is assumed to be $1^{\circ}$ at most and if the sky is covered relatively uniformly, the supports of the functions $\tilde{f}({\bf l})$ and $\tilde{h_{\pm}}({\bf l})$ are narrow with a support on a very small range of $\Delta l$. Combining these two facts together it is evident that we can still work in the flat-sky approximation in most cases of interest.
  
We limit the following discussion to gaussian beams
but similar calculations can be readily done in the case of other beam 
shapes (at least numerically).
\begin{table}[h!b!p!]
\begin{tabular}[c]{|c|c|c|}
\hline
~parameter~&~beam 1~&~beam 2~\\
\hline
~$\sigma_{x}$~& $ \sigma(1+\mu)(1+e) $ & $\sigma(1-\mu)(1+e)$ \\
\hline
~$\sigma_{y}$~& $ \sigma(1+\mu)(1-e) $ & $\sigma(1-\mu)(1-e)$ \\
\hline
~$y$~& $\frac{(l\sigma)^{2}}{2}(1+\mu)^{2}(1+e^{2})$ & $\frac{(l\sigma)^{2}}{2}(1-\mu)^{2}(1+e^{2})$ \\
\hline
~$z$~& $(l\sigma)^{2}(1+\mu)^{2}e$ & $(l\sigma)^{2}(1-\mu)^{2}e$ \\
\hline
\end{tabular}
\caption{Definitions of the parameters associated with the dual-beam experiment.} 
\end{table}
The corrections to the underlying power spectra are defined as follows
\begin{eqnarray}
\Delta C^{Z}_{l}&\equiv &
C'^{Z}_{l}(\sigma,g_{i},\mu_{i},e_{i},\rho_{i},\varepsilon)-C'^{Z}_{l}(\sigma,0)\nonumber\\
Z &\in &\{TT,TE,EE,BB,TB,EB\}
\end{eqnarray}
which are functions of the small parameters $g$, ${\bf\rho}$, $\mu$,
$e$, $\varepsilon$ and the underlying power spectra, where here $Z=TT,TE,EE,BB,TB$ 
and $EB$, and as we see from Eqs. (21) and (24), the new power spectra are, in principle,
combinations of all the underlying power spectra. 
Also, it is worth mentioning here the coupling
between the various systematic effects as far as higher order corrections are
concerned. While Eqs.(21) and (12) are presented in terms of 
formally-exact, well-known
functions, for calculations the infinite series (Eq. 12) 
must be truncated, and depending on the degree of asymmetry and the physical
scale in question (tantamount to the multipole number $l$), 
the number of terms in the series considered determines the accuracy of the 
calculation. The scalings of the leading order terms of the effects
considered here are given in Tables III-VI assuming the underlying sky is unpolarized
(except for the effect of rotation, to be discussed in the next section).
In obtaining these expressions we also assumed the scanning strategy is statistically isotropic, 
a reasonable assumption that significantly simplify the scaling relations we obtain.
We have used the following definitions
\begin{eqnarray}
f_{1}&=&\frac{1}{2}|\tilde{h}_{+}(-1,0)|^{2}\nonumber\\
f_{2}&=&\frac{1}{2}|\tilde{h}_{+}(-1,-1)|^{2}+\frac{1}{2}|\tilde{h}_{+}(-1,1)|^{2}\nonumber\\
f_{3}&=&\frac{1}{2}\langle\tilde{f}(0,1)\tilde{h}_{-}^{*}(1,-1)\rangle
\end{eqnarray}
where the functions $f(m,n)$ and $h_{\pm}(m,n)$ are defined in Eq.(20).
Since the leading orders of both $J_{n}(z)$ and $I_{n}(z)$ 
are $\propto z^{n}$ it is clear that terminating the series at 
some order $n$ is equivalent to a
power series expansion exact up to this order and not higher. 
Therefore, the calculation
described here actually amounts to a power series expansion 
in the most general case, but simplifies considerably when either the dipole 
or quadrupole effects can be ignored. 
Tables V and VI show the scaling relations in the case of ideal scanning strategy 
corresponding to Tables III and IV, respectively. These are the irreducible signals 
that persist irrespective of the scanning strategy. 
One further remark is in order here: the coupling
between the various effects is important when the parameters $l^{2}\sigma^{2}\mu$, $l\rho$
and $l^{2}\sigma^{2}e$ are not negligibly small compared to $1$; in this case higher order
corrections are required and the cross terms that include the coupling between the effects
cannot be ignored, but this is seldom the case. 
For given parameters $\mu$, $e$ and $\rho$, there is a `critical' value of
$l$ beyond which higher order terms become important and in this range our calculation
may be particularly useful. 
For $\sigma\approx 1^{\circ}$ and 
$e$, $\mu$ and $\rho/\sigma$ $\approx 0.1$ 
higher order effects become important at $l\approx 1000$ which is tantalizingly close 
to the scale at which the B-mode signal from lensing peaks (Zaldarriaga \& Seljak [19], Hu [20]).
However, it happens deep beyond the beam dilution scale, $l\approx 200$.
Only when $\rho$ is not very small compared to $\sigma$ or $e$ and $\mu$ are not 
small compared to unity do these higher order corrections contribute to the spurious polarization.

\subsection{Rotations}

We now discuss the rotation error ignored in the above treatment. 
Overall rotation of the two beams 
(cross-polarization) mixes E and B-modes and even induces T-B correlation. 
Non-orthogonality of the beams
can be described in two different ways. If only one of the two
detectors is miss-oriented by $\delta$, there is induced E-B mixing.
This is the approach taken by Ponthieu [10]. If the two detectors are 
disoriented, one by $+\delta/2$, the other one by $-\delta/2$, the Q-
U mixing induced by the first detector is compensated by that of the
second detector and there is no E-B mixing. In the following, we
focus on the latter when we refer to {\it non-orthogonality} of the
beams. 
We here derive the expected signal. For (overall) rotation error $\varepsilon$ and 
non-orthogonality measured by $\delta$ (which we split between the two beams; 
one is missoriented by $\delta/2$ and the other by $-\delta/2$)  
\begin{eqnarray}
& &\sum_{j}(A^{T}d_{j})({\bf p})=\parallel M\parallel{\bf I}
\end{eqnarray}
where ${\bf I}\equiv(T,Q-iU,Q+iU)$ and
\begin{eqnarray}
&&\parallel M\parallel\equiv\nonumber\\
&&\left(\begin{array}{c c c}
1 & \frac{1}{2}\langle e^{2i\alpha}\rangle e^{2i\varepsilon+i\delta}& \frac{1}{2}\langle e^{-2i\alpha}\rangle e^{-2i\varepsilon-i\delta}\\
\frac{1}{2}\langle e^{2i\alpha}\rangle & \frac{1}{4}\langle e^{4i\alpha}\rangle e^{2i\varepsilon+i\delta}& 
\frac{1}{4}e^{-2i\varepsilon-i\delta}\\
\frac{1}{2}\langle e^{-2i\alpha}\rangle & \frac{1}{4}e^{2i\varepsilon+i\delta} & 
\frac{1}{4}\langle e^{-4i\alpha}\rangle e^{-2i\varepsilon-i\delta}
\end{array}\right)
\end{eqnarray}
and upon using Eq.(16), we obtain
\begin{eqnarray}
&&\left(\begin{array}{c}
T'\\
Q'-iU'\\
U'+iU'\\
\end{array}\right)=\nonumber\\
&&\left(\begin{array}{c}
T+\frac{i}{2}\sin\delta\sum_{\pm}F_{\pm}(Q\mp iU)e^{\pm 2i\varepsilon}\langle e^{\pm 2i\alpha}\rangle\\
(Q-iU)\cos\delta e^{2i\varepsilon}\\
(Q+iU)\cos\delta e^{-2i\varepsilon}
\end{array}\right)
\end{eqnarray}
where $F_{\pm}\equiv\mp 1$.
Employing Eqs.(3)
\begin{eqnarray}
\tilde{Q}_{{\bf l}}\pm i\tilde{U}_{{\bf l}}=(E_{{\bf l}}\pm iB_{{\bf l}})e^{\mp 2i\phi_{l}}
\end{eqnarray}
we obtain to leading order
\begin{eqnarray}
T'_{{\bf l}}&\approx &T_{{\bf l}}\nonumber\\
E'_{{\bf l}}&\approx &E_{{\bf l}}-2\varepsilon B_{{\bf l}}\nonumber\\
B'_{{\bf l}}&\approx &B_{{\bf l}}+2\varepsilon E_{{\bf l}} 
\end{eqnarray}
and therefore
\begin{eqnarray}
&&\Delta C_{l}^{T'}\approx 0\nonumber\\
&&\Delta C_{l}^{T'E'}\approx 0\nonumber\\
&&\Delta C_{l}^{E'}\approx 0\nonumber\\
&&\Delta C_{l}^{B'}\approx 4\varepsilon^{2}C_{l}^{E}\nonumber\\
&&\Delta C_{l}^{T'B'}\approx  2\varepsilon C_{l}^{TE}\nonumber\\
&&\Delta C_{l}^{E'B'}\approx 2\varepsilon C_{l}^{E}
\end{eqnarray}
i.e. the leakage is a small fraction of $C^{E}_{l}$ but can still be a significant 
B-mode contaminant.

\subsection{Monitoring the Contamination}

A potentially useful diagnostic is the $T-B$ cross
correlation. As mentioned above, 
this correlation function vanishes in the standard model. 
The spurious effects discussed in this paper cause T to leak to both E and B, 
and therefore the 
correlations $C_{l}^{TB}$ and $C_{l}^{EB}$ do not generally vanish 
(see Eqs. 24 and Table IV). Furthermore, as we will see below, beam 
rotation can also induce $C_{l}^{TB}$ and $C_{l}^{EB}$ because 
rotating the telescope by $\varepsilon$ is indistinguishable 
from rotation of the polarization plane by $-\varepsilon$.
As a result, power leaks from T-E to T-B and from E-E to E-B.
A nonvanishing $C^{TB}_{l}$ may be attributed to 
an imperfect removal of the spurious polarization 
signals discussed in this paper. 
However, beam systematics are not the exclusive generating 
mechanisms of $C_{l}^{TB}$ and $C_{l}^{EB}$. A few {\it physical} 
sources of parity-violating correlations were already discussed in 
the literature; parity-violating terms in the lagrangians of the 
electromagnetic and gravitational sectors
(e.g., Lue, Wang \& Kamionkowski [21], Liu, Lee \& Ng [22], 
Feng et al. [23], Saito, Ichiki \& Taruya [24], Xia et al. [25], 
Komatsu et al. [26]), 
Faraday rotation at last scattering (Kosowsky \& Loeb [27]), 
and hypothetical primordial helical magnetic fields (Caprini, Durrer \& 
Kahniashvili [28]). The systematic T-B and E-B correlations may 
interfere with this exotic physics and a careful analysis 
of these correlations in required in the presence of beam systematics 
(Shimon \& Keating [29]). 

\begin{table}[c]
\begin{tabular}{|c|c|c|c|c|c|}
\hline
~effect~&~parameter~&~$\Delta C_{l}^{TE}$~&~$\Delta C_{l}^{E}$~&~$\Delta C_{l}^{B}$~\\
\hline
~gain~& $g$ & 0 & $g^{2}f_{1}\star C_{l}^{T}$ & $g^{2}f_{1}\star C_{l}^{T}$\\
\hline
~monopole~& $\mu$ & 0 & $4\mu^{2}(l\sigma)^{4}f_{1}\star C_{l}^{T}$ & $4\mu^{2}(l\sigma)^{4}f_{1}\star C_{l}^{T}$\\
\hline
~pointing~& $\rho$ & $\frac{1}{2}c_{\theta}C_{l}^{T}[1+J_{0}(l\rho)]J_{2}(l\rho)$ & $c_{\theta}^{2}C_{l}^{T}J_{2}^{2}(l\rho)$ & $s_{\theta}^{2}C_{l}^{T}J_{2}^{2}(l\rho)$\\
            &     & $-c_{\theta}J_{1}^{2}(l\rho)C_{l}^{T}\star f_{3}$ & $+J_{1}^{2}(l\rho)C_{l}^{T}\star f_{2}$&$-J_{1}^{2}(l\rho)C_{l}^{T}\star f_{2}$\\
\hline
~quadrupole~& e & $-I_{0}(z)I_{1}(z)c_{\psi}C_{l}^{T}$ & 
$I_{1}^{2}(z)c_{\psi}^{2}C_{l}^{T}$ & $I_{1}^{2}(z)s_{\psi}^{2}C_{l}^{T}$ \\
\hline
~rotation~& $\varepsilon$ & $0$ & $4\varepsilon^{2}C_{l}^{B}$ & $4\varepsilon^{2}C_{l}^{E}$\\
\hline
\end{tabular}
\caption{The leading order contributions of the systematic effects to 
the power spectra $C_{l}^{TE}$, $C_{l}^{E}$ and $C_{l}^{B}$ 
assuming the underlying sky is not polarized (except 
for the {\it rotation} signal where we assume the E-, and B-mode signals are present) 
and general sky scanning.} 
\end{table}

\begin{table}[c]
\begin{tabular}[c]{|c|c|c|c|}
\hline
~effect~&~parameter~&~$\Delta C_{l}^{TB}$~&~$\Delta C_{l}^{EB}$~\\
\hline
~gain~& $g$ & 0 & 0 \\
\hline
~monopole~& $\mu$ & 0 & 0\\
\hline
~pointing~& $\rho$ & $\frac{1}{2}J_{2}(l\rho)[1+J_{0}(l\rho)]s_{\theta}C_{l}^{T}$ & $s_{\theta}c_{\theta}J_{2}^{2}(l\rho)C_{l}^{T}$ \\
 &  & $+s_{\theta}J_{1}^{2}(l\rho)C_{l}^{T}\star f_{3}$ & \\
\hline
~quadrupole~& $e$ & $-I_{0}(z)I_{1}(z)s_{\psi}C_{l}^{T}$ & 
$I_{1}^{2}(z)s_{\psi}c_{\psi}C_{l}^{T}$ \\
\hline
~rotation~& $\varepsilon $ & $2\varepsilon C_{l}^{TE}$ & $2\varepsilon C_{l}^{E}$\\
\hline
\end{tabular}
\caption{The contribution of the systematic effects to 
the power spectra $C_{l}^{TB}$, $C_{l}^{EB}$ 
assuming the underlying sky is not polarized (except 
for the {\it rotation} signal when we assume E-, and B-mode polarization are present) 
and general sky scanning.}
\end{table}

\begin{table}[c]
\begin{tabular}{|c|c|c|c|c|c|}
\hline
~effect~&~parameter~&~$\Delta C_{l}^{TE}$~&~$\Delta C_{l}^{E}$~&~$\Delta C_{l}^{B}$~\\
\hline
~gain~& $g$ & 0 & 0 & 0 \\
\hline
~monopole~& $\mu$ & 0 & 0 & 0\\
\hline
~pointing~& $\rho$ & $
\frac{1}{2}(1+J_{0}(l\rho))J_{2}(l\rho)c_{\theta}C_{l}^{T}$ & $J_{2}^{2}(l\rho)c_{\theta}^{2}C_{l}^{T}$ & $J_{2}^{2}(l\rho)s_{\theta}^{2}C_{l}^{T}$\\
\hline
~quadrupole~& e & $-I_{0}(z)I_{1}(z)c_{\psi}C_{l}^{T}$ & 
$I_{1}^{2}(z)c_{\psi}^{2}C_{l}^{T}$ & $I_{1}^{2}(z)s_{\psi}^{2}C_{l}^{T}$ \\
\hline
~rotation~& $\varepsilon$ & $0$ & $4\varepsilon^{2}C_{l}^{B}$ & $4\varepsilon^{2}C_{l}^{E}$\\
\hline
\end{tabular}
\caption{The contribution of the systematic effects to 
the power spectra $C_{l}^{TE}$, $C_{l}^{E}$ and $C_{l}^{B}$ 
assuming the underlying sky is not polarized (except 
for the {\it rotation} signal where we assume the E-, and B-mode signals are present) 
and ideal sky scanning.}
\end{table}
\begin{table}[c]
\begin{tabular}[c]{|c|c|c|c|}
\hline
~effect~&~parameter~&~$\Delta C_{l}^{TB}$~&~$\Delta C_{l}^{EB}$~\\
\hline
~gain~& $g$ & 0 & 0 \\
\hline
~monopole~& $\mu$ & 0 & 0\\
\hline
~pointing~& $\rho$ & $\frac{1}{2}[1+J_{0}(l\rho)]J_{2}(l\rho)s_{\theta}C_{l}^{T}$ & $s_{\theta}c_{\theta}J_{2}^{2}(l\rho)C_{l}^{T}$\\
\hline
~quadrupole~& $e$ & $-I_{0}(z)I_{1}(z)s_{\psi}C_{l}^{T}$ & 
$I_{1}^{2}(z)s_{\psi}c_{\psi}C_{l}^{T}$ \\
\hline
~rotation~& $\varepsilon $ & $2\varepsilon C_{l}^{TE}$ & $2\varepsilon C_{l}^{E}$\\
\hline
\end{tabular}
\caption{The contribution of the systematic effects to 
the power spectra $C_{l}^{TB}$, $C_{l}^{EB}$ 
assuming the underlying sky is not polarized (except 
for the {\it rotation} signal when we assume E-, and B-mode polarization are present) 
and ideal sky scanning.}
\end{table}

\section{Numerical Results}

The Stokes $Q$ parameters associated with the polarization due to the
optical imperfections discussed in this paper are shown in Figure 2.
\begin{figure*}
\includegraphics[width=18.5cm]{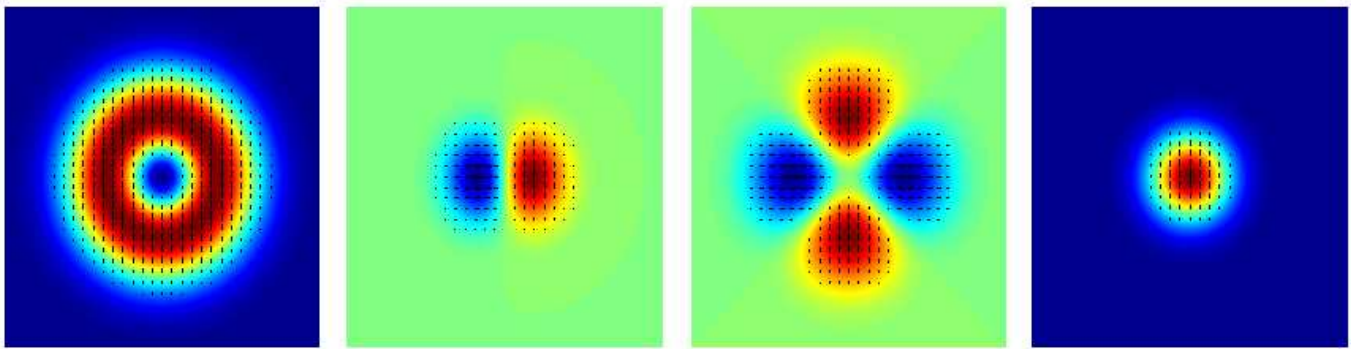}
\caption{An illustration of the monopole, dipole, quadrupole and gain
effects; Q parameter only is depicted.}
\end{figure*}
We have calculated all possible six power spectra (Eqs. 24 and 33)
for two different average beam widths FWHM $56'$ and $5'$.
Figures 3-8 depict beam rotation, differential pointing and 
differential ellipticities effects on the $B$-mode power spectrum 
for average beamwidths of
$56'$ and $5'$, respectively. The dot-dashed lines refer to the
inflation-induced $B$-mode from primordial gravitational 
waves with tensor to scalar ratios of 
$T/S=10^{-1}$, $10^{-2}$, $10^{-3}$ and $10^{-4}$, 
respectively, where we have used the definition used in
CAMB for the tensor-to-scalar ratio.
For the plots of the second order differential pointing effect the specified 
pointing error $\rho$ refers to one of the beams (we left the angle $\theta$ 
of this beam a free parameter), while the other 
beam has been assumed to have no pointing error . The quadrupole effect was calculated 
assuming the two beams have the {\it same} specified ellipticities, $|e|$, and here 
we left the angles that the polarization axes make with the major axes of the two beams, 
$\psi_{1}$ and $\psi_{2}$, as free parameters. 
\begin{figure*}
\includegraphics[width=10.5cm]{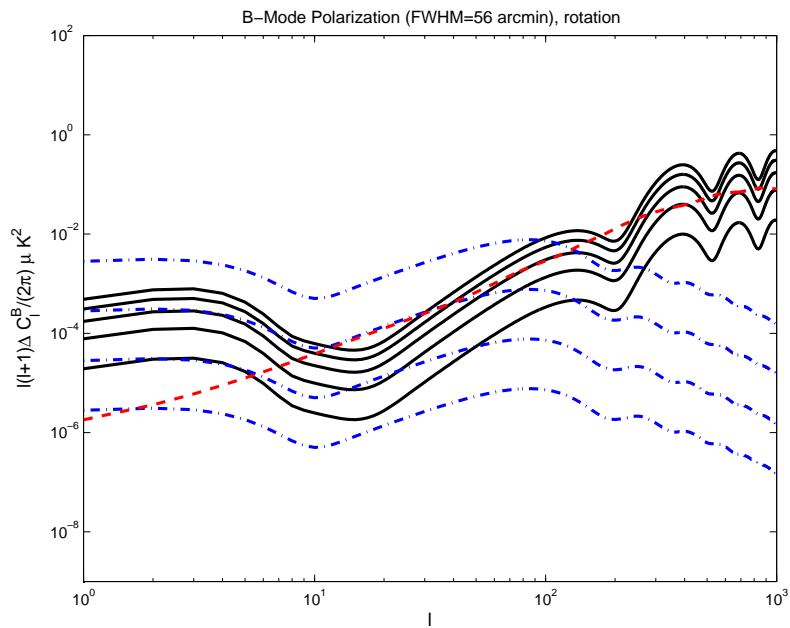}
\caption{The contribution of differential rotation to the $B$-mode
power spectrum ($56'$ average beamwidth). Shown are the effects for 
$\varepsilon=0.01,\ 0.02,\ 0.03,\ 0.04$ and $0.05$ of a radian. 
For comparison, the dot-dashed curves
refer to the contribution from primordial gravitational waves with tensor
to scalar ratios $T/S=10^{-1}$, $10^{-2}$, $10^{-3}$ and $10^{-4}$. 
The dashed curve is the $B$-mode polarization produced by gravitational lensing by the large scale structure.} 
\end{figure*}
\begin{figure*}
\includegraphics[width=10.5cm]{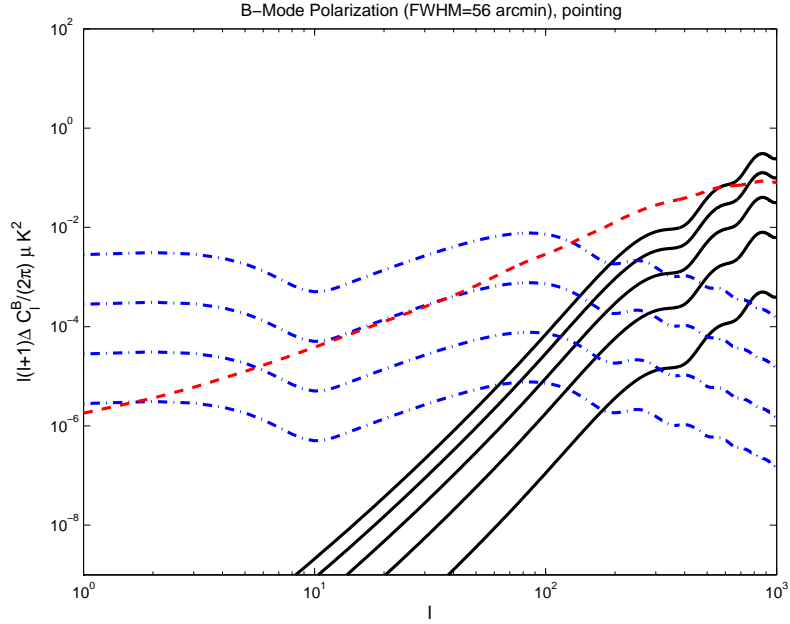}
\caption{The contribution of differential pointing to the $B$-mode
power spectrum ($56'$ average beamwidth). 
The values shown should be multiplied by $s_{\theta}^{2}$ (Table III).
In case $\theta=0^{\circ}$ all the spurious polarization is in the E-mode.
For a given $\rho$, the maximum B-mode is obtained at $\theta=45^{\circ}$. 
Shown are the effects for 
$\rho=0.01,\ 0.02,\ 0.03,\ 0.04$ and $0.05$. 
For comparison, the dot-dashed curves
refer to the contribution from primordial gravitational waves with tensor
to scalar ratios $T/S=10^{-1}$, $10^{-2}$, $10^{-3}$ and $10^{-4}$. 
The dashed curve is the $B$-mode polarization produced by gravitational lensing by the large scale structure.}
\end{figure*}
\begin{figure*}
\includegraphics[width=10.5cm]{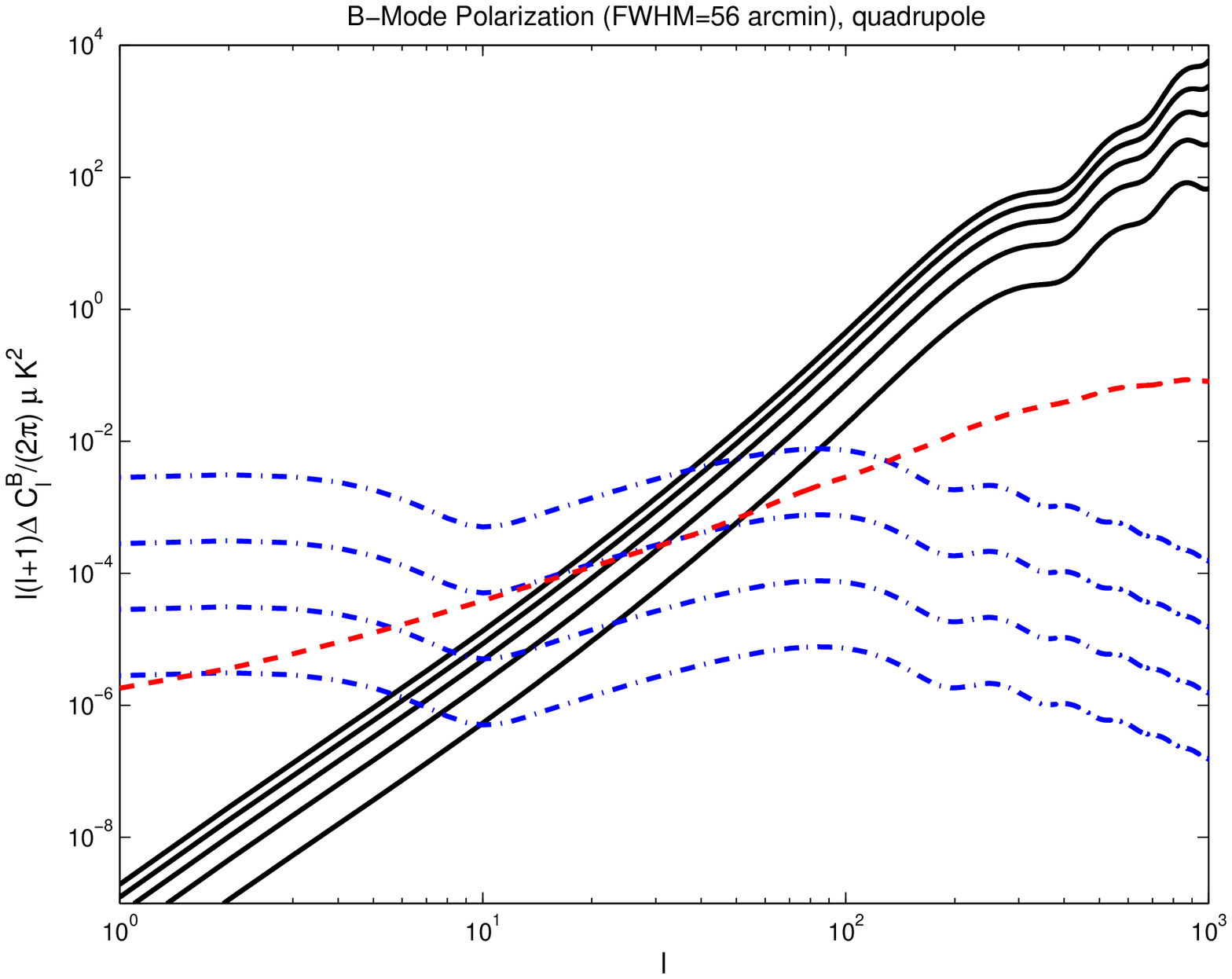}
\caption{The contribution of the differential beam ellipticity 
(`quadrupole' effect) to the $B$-mode
power spectrum ($56'$ average beamwidth). 
The values shown should be multiplied by $s_{\psi}^{2}$ (see Table III).
Shown are the effects for $e=0.01,\ 0.02,\ 0.03,\ 0.04$ and $0.05$. 
For comparison, the dot-dashed curves
refer to the contribution from primordial gravitational waves with tensor
to scalar ratios $T/S=10^{-1}$, $10^{-2}$, $10^{-3}$ and $10^{-4}$. 
The dashed curve is the $B$-mode polarization produced by gravitational lensing by the large scale structure.}
\end{figure*}
\begin{figure*}
\includegraphics[width=10.5cm]{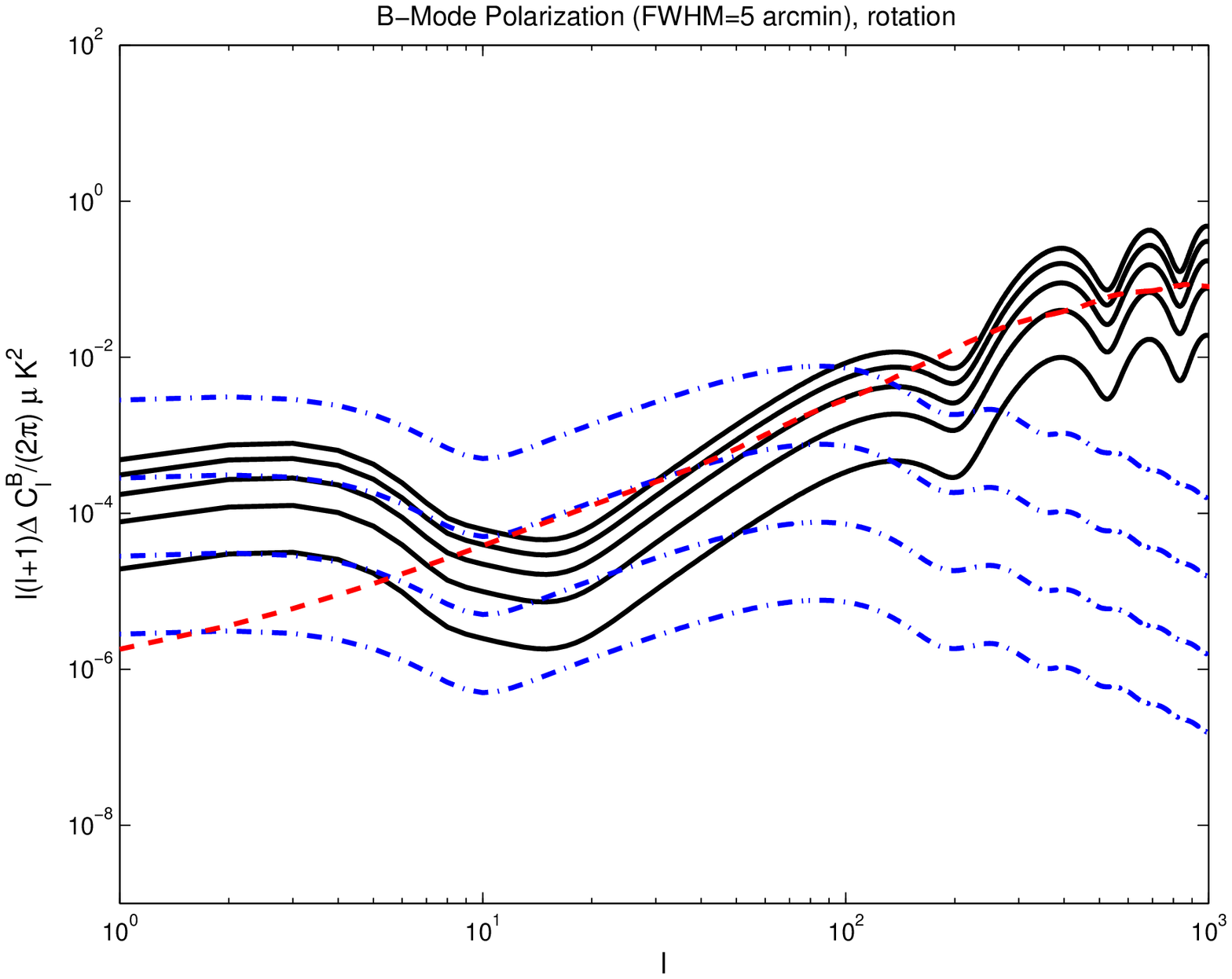}
\caption{The contribution of differential rotation to the $B$-mode
power spectrum ($5'$ average beamwidth). Shown are the effects for 
$\varepsilon=0.01,\ 0.02,\ 0.03,\ 0.04$ and $0.05$ of a radian. 
For comparison, the dot-dashed curves 
refer to the contribution from primordial gravitational waves with tensor
to scalar ratios $T/S=10^{-1}$, $10^{-2}$, $10^{-3}$ and $10^{-4}$. 
The dashed curve is the $B$-mode polarization produced by gravitational lensing by the large scale structure.}
\end{figure*}
\begin{figure*}
\includegraphics[width=10.5cm]{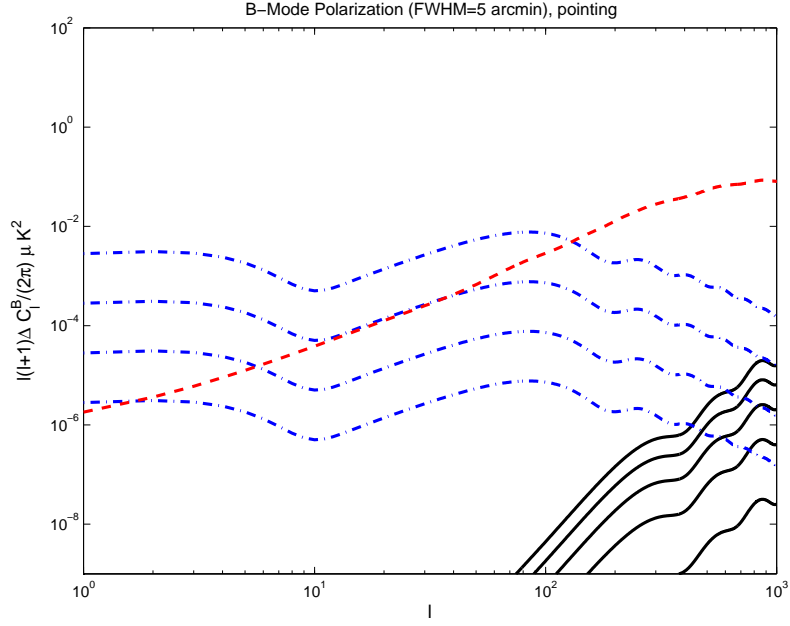}
\caption{The contribution of differential pointing to the $B$-mode
power spectrum ($5'$ average beamwidth). 
The values shown should be multiplied by $s_{\theta}^{2}$. 
Shown are the effects for 
$\rho=0.01,\ 0.02,\ 0.03,\ 0.04$ and $0.05$. 
For comparison, the dot-dashed curves
refer to the contribution from primordial gravitational waves with tensor
to scalar ratios $T/S=10^{-1}$, $10^{-2}$, $10^{-3}$ and $10^{-4}$. 
The dashed curve is the $B$-mode polarization produced by gravitational lensing by the large scale structure.}
\end{figure*}
\begin{figure*}
\includegraphics[width=10.5cm]{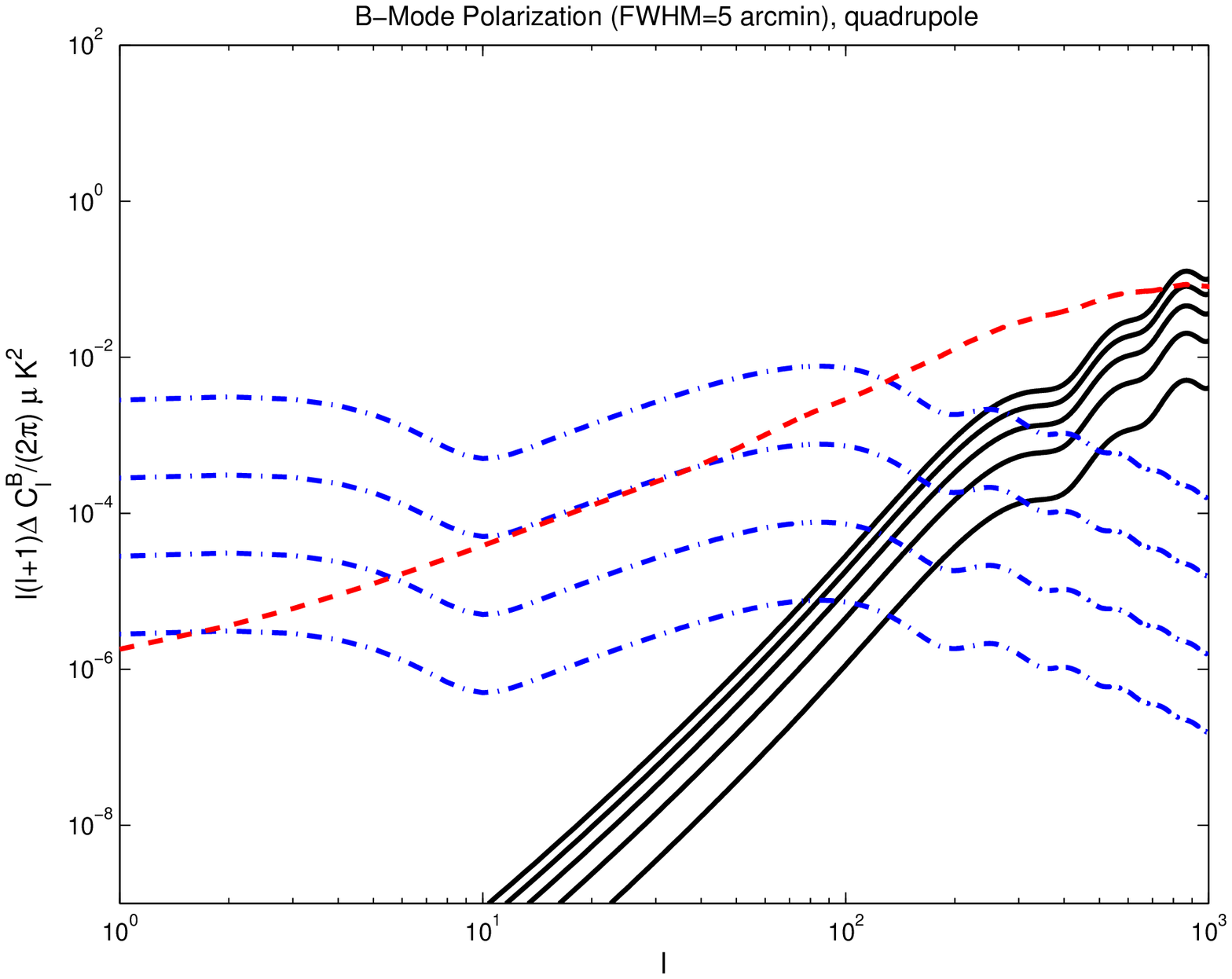}
\caption{The contribution of the differential beam ellipticity 
(`quadrupole' effect) to the $B$-mode power spectrum 
($5'$ average beamwidth). 
The values shown should be multiplied by $s_{\psi}^{2}$. 
Shown are the effects for 
$e=0.01,\ 0.02,\ 0.03,\ 0.04$ and $0.05$ of a radian. 
For comparison, the dot-dashed curves
refer to the contribution from primordial gravitational waves with tensor
to scalar ratios $T/S=10^{-1}$, $10^{-2}$, $10^{-3}$ and $10^{-4}$. 
The dashed curve is the $B$-mode polarization produced by gravitational lensing by the large scale structure.}
\end{figure*}
We have defined $c_{\psi}\equiv\cos(2\psi_{1})-\cos(2\psi_{2})$,
$s_{\psi}\equiv\sin(2\psi_{1})-\sin(2\psi_{2})$ and $c_{\theta}\equiv\cos 2\theta$, 
$s_{\theta}\equiv\sin 2\theta$. 
Note that in the case considered here, that both beams have the same ellipticity $|e|$, 
we obtain, as expected, that there in no spurious polarization when 
$\psi_{1}=\psi_{2}$. 
Similarly, had we assumed both beams have the same pointing $\rho$ 
we would have obtained no spurious polarization if $\theta_{1}=\theta_{2}$ 
in the case of the ideal scanning, where second order pointing is the 
leading order contribution (not so in the case of non-ideal scanning, 
where the leading order is the dipole effect).

As expected, the larger the beamwidths 
the larger the angular scales on which
the systematics peak. The systematics are calculated 
for the small parameters;
$g$, $\mu$ and $e=$ $1,2,3,4$ and $5$ \% of the mean beam 
and $\rho$ and $\varepsilon$ the same fractions of the mean beam and a radian, 
respectively. The scaling relations of these effects are given in Tables III-IV 
(general scanning strategy) and V-VI (ideal scanning strategy). 
As explained in
the last section, the higher order corrections (beyond the leading order) 
are important on some scale, typically $l\approx\sigma^{-1}\mu^{-\frac{1}{2}}$ 
which for $\sigma\approx 1^{\circ}$ and $e$ and $\mu$ on the 1\% level corresponds to 
$l\approx 2000$. On this scale, the $B$-mode from the CMB lensed by the large 
scale structure (e.g. Zaldarriaga \& Seljak [19], Hu [20]) 
is non-negligible (the dashed lines in Figures 3 through 8). 
However a beam that size is insensitive to features on 
scales of $l\approx 2000$.
It is also important to mention here that the effect due to rotation 
(Figures 3 and 6) closely follows the $C_{l}^{E}$ shape and merely reflects 
the leakage of $E$ to $B$ due to rotation (Eq. 33). One more remark
should be made regarding the coupling of the differential pointing  
and quadrupole effects mentioned above. 
As seen from Eqs. (12) the pointing error and ellipticity are
coupled through products of the cylindrical and modified Bessel functions. In
practice however the coupling is small since the pointing error parameter 
$l\rho$ is small. In this case, since the leading term of $J_{n}(x)$ is
$\propto x^{n}$ it is safe to 
consider only the leading terms in the infinite sum in 
Eq. (12), this approximation 
considerably simplifies numerical calculations.
In calculating the plots we assumed perfect scanning strategy so the monopole and 
gain identically vanish as expected and as can be easily verified from Eqs. (12), 
(21) and the relations in the Appendix. Again, it is important to note that the power spectra for the 
second order pointing error (Figures 4 and 7) and quadrupole (Figures 5 and 8) 
should be multiplied by the functions $c_{\theta}$ and $s_{\theta}$, and 
$c_{\psi}$ and $s_{\psi}$, respectively, as described in 
Tables III, IV, V \& VI. These functions can vanish. For instance, the effect of ellipticity 
will not contribute to the B-mode power spectra if the polarization sensitive 
axes are parallel to one of the ellipse principal axes ($\psi=0,\pi/2$), there is no shear 
of the field in this highly symmetric case and all the spurious polarization is in the 
E-mode as can be seen from Table V. This is no longer the case with non-ideal scanning strategy 
(Table III); in this case the first order dipole do not depend on $\theta$. 
Also, if $\psi_{1}=\psi_{2}$ 
there will be no induced polarization by differential ellipticity, not in the E-, nor 
in the B-mode.
Similarly, for the pointing error, if $\theta=0$, 
all the spurious polarization will contaminate the E-mode 
(if the scanning strategy is ideal).
The power spectrum associated with the underlying sky was 
calculated by CAMB using cosmological parameters consistent with WMAP 
(Spergel et al. [30]). 
We ignore gravitational lensing and the tensor contribution 
to the underlying sky, but we do show them in the figures for reference. 
We stress again
that nonvanishing $C'^{TB}_{l}$ and $C'^{EB}_{l}$ may indicate an
incomplete removal of the spurious polarization signals discussed here.
The relevant expressions are given in Tables IV \& VI.  
Figures 9 to 11 show that $C'^{TB}_{l}$ peaks at the few $\mu$K level 
(all these values are for `maximum shear', i.e. $\theta=45^{\circ}$ or $\psi=45^{\circ}$), 
and so constitute only upper limits for the given parameters $\rho$ and $e$.
\begin{figure*}
\includegraphics[width=10.5cm]{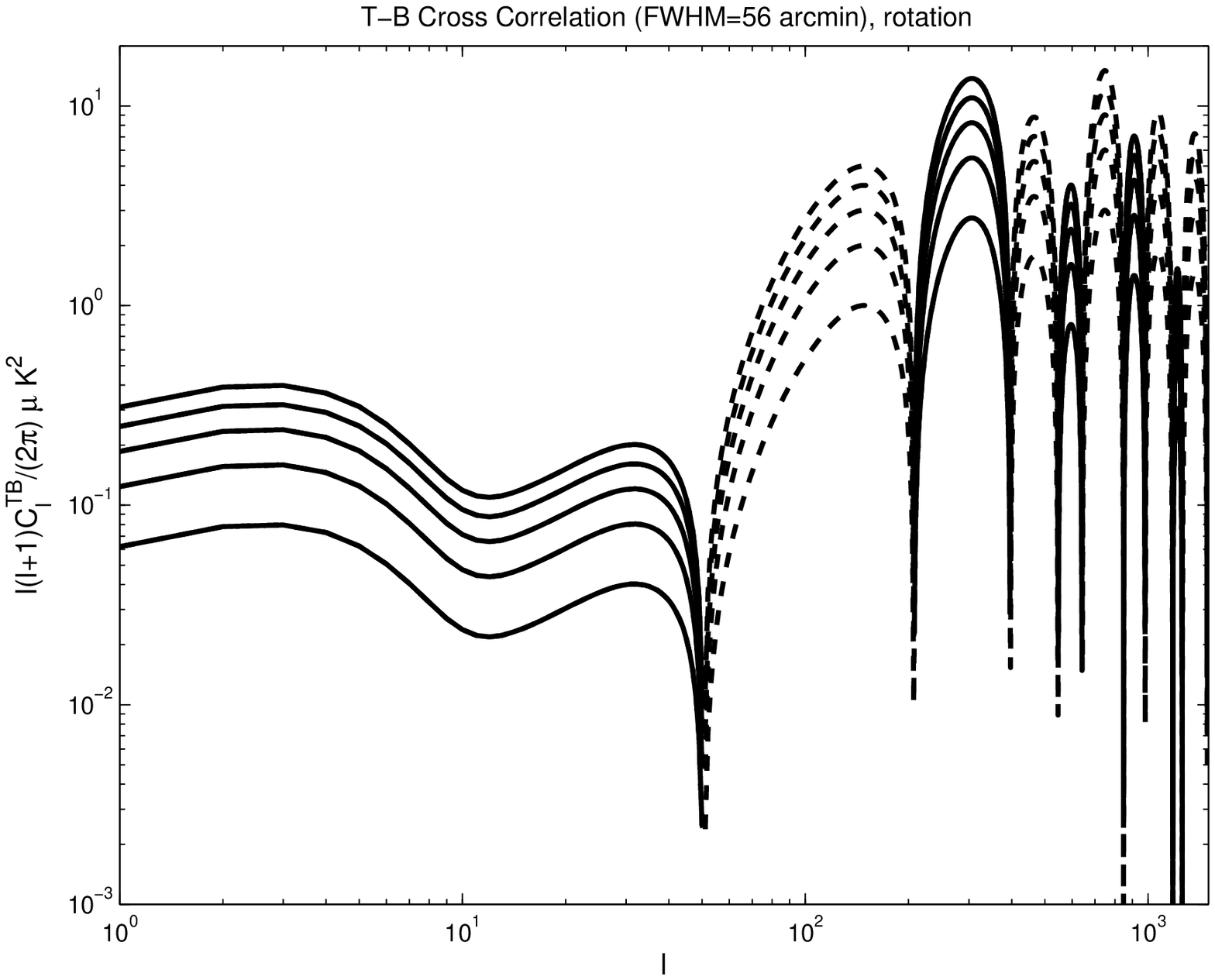}
\caption{The contribution of differential rotation to the $T-B$ cross correlation 
($56'$ average beamwidth). Shown are the signals for 
$\varepsilon=0.01,\ 0.02,\ 0.03,\ 0.04$ and $0.05$ of a radian. 
Dashed lines correspond to negative values of the cross-correlation 
induced by the underlying T-E cross correlation.}
\end{figure*}
\begin{figure*}
\includegraphics[width=10.5cm]{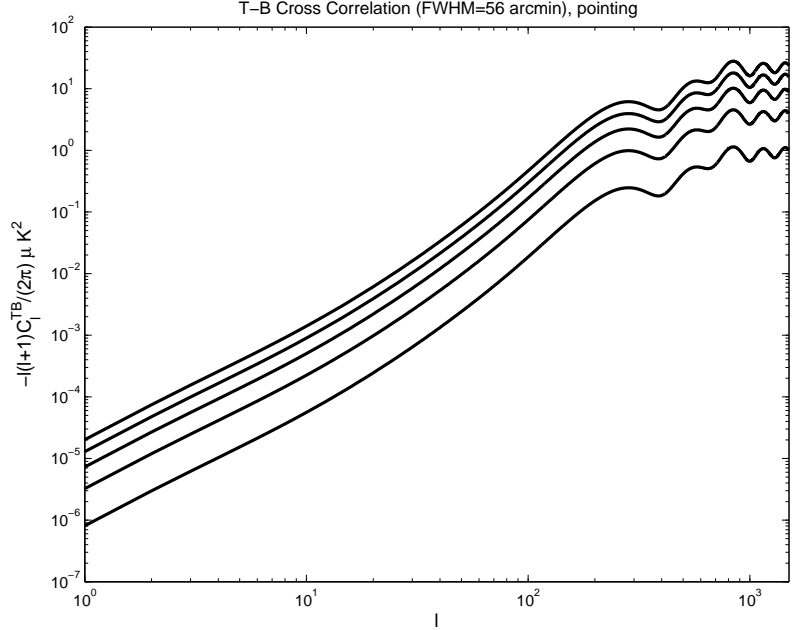}
\caption{The contribution of differential pointing to the $T-B$ cross correlation 
($56'$ average beamwidth). The values shown should be multiplied by 
$s_{\theta}$. Shown are the effects for $\rho=0.01,\ 0.02,\ 0.03,\ 0.04$ and $0.05$.}
\end{figure*}
\begin{figure*}
\includegraphics[width=10.5cm]{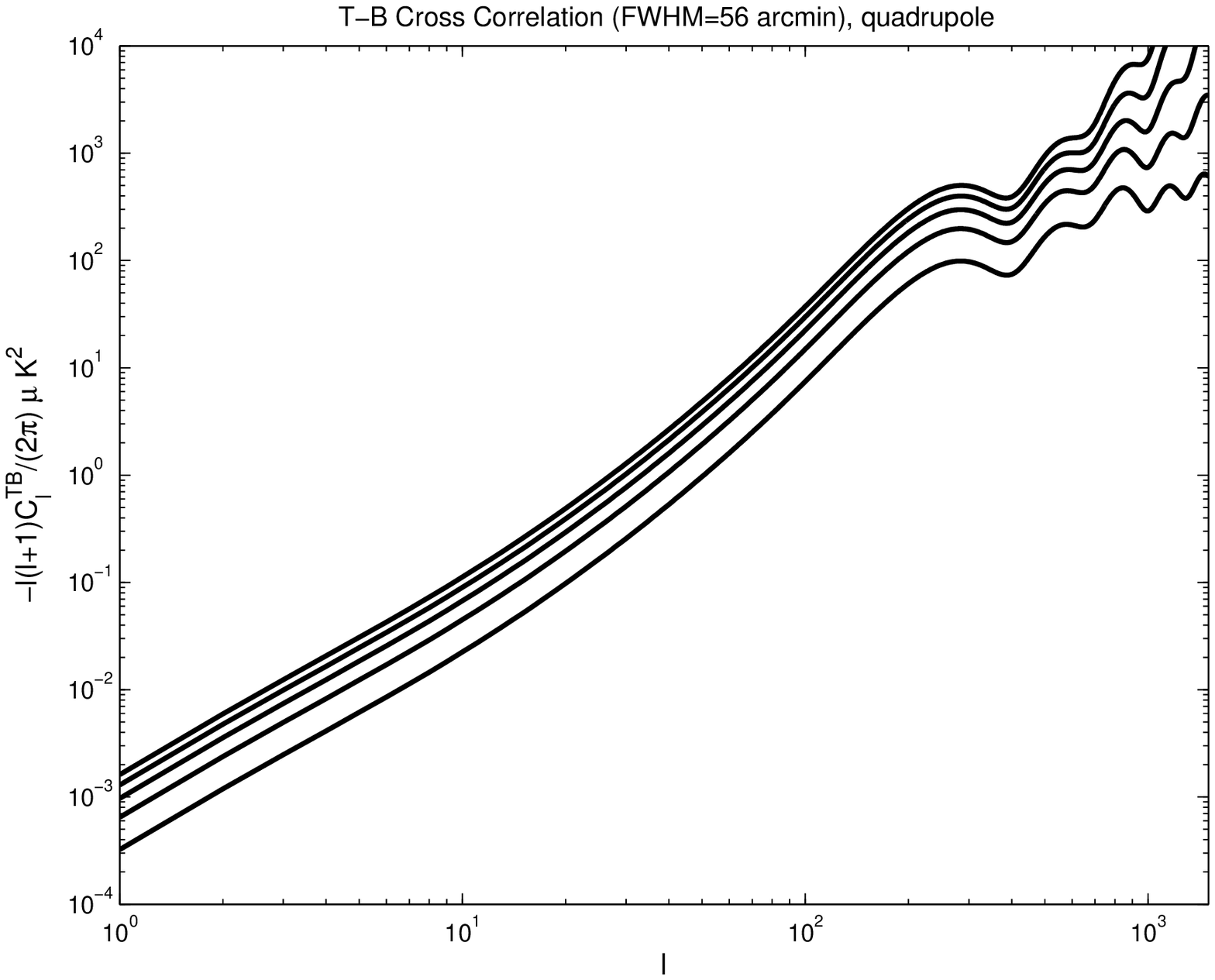}
\caption{The contribution of the `quadrupole' to the $T-B$ cross correlation 
($56'$ average beamwidth). 
The values shown should be multiplied by $s_{\psi}$.
Shown are the effects for 
$e=0.01,\ 0.02,\ 0.03,\ 0.04$ and $0.05$.}
\end{figure*}
Only the contributions from (second order) differential pointing, differential ellipticity 
and differential rotation  for the $TB$ cross-correlation are 
displayed since the monopole and gain contributions 
identically vanish for an ideal scanning strategy. 
One more aspect of this analysis is 
the higher order corrections of the effects studied in this paper. 
We find that these corrections have little effect even on the smallest scales 
(where the gradients are large) in the ideal cases we studied. These effects become 
relatively large on scales much smaller than the mean beam size where beam dilution 
is significant. We conclude that at least for nearly ideal scanning strategy these 
higher order corrections can be safely neglected.

\section {Discussion}

The systematics discussed here, 
more than simply affecting the peak of the B-mode power 
spectrum at $l\approx 100$ are likely to impact the 
polarization signal due to lensing at $l\approx$ few hundreds as shown in Figures
3 to 8, at least for small beams. 
A full analysis of the effect of gravitational waves on CMB 
B-mode polarization is underway, and requires an assessment of the effect of gravitational 
lensing. The results presented in this paper show that 
the beam systematics are likely to further complicate this 
process both directly (with a residual effect on degree scales) 
and indirectly (by mimicking the effect of lensing).
While the lensing signal can be partially extracted by invoking 
optimal estimators that use the non-gaussianity of the lensed anisotropy 
and polarization (Hu \& Okamoto [31]), the systematics discussed here cannot be removed by this method 
because they inherit the statistics of the underlying sky, which is gaussian in 
the standard model.    
The lensing-induced polarization is a direct probe of structure formation
processes and is sensitive to few cosmological parameters, most notably the neutrino mass.
An accurate analysis should take the beam effects discussed here into account. 
Our calculation demonstrates that
at most, these systematics are on the level of few tens to few hundredths of nK
and their significance as contaminants of the primordial B-mode 
depends on the tensor to scalar ratio. Comparing our results 
to the results of Ponthieu [10] it is evident that the spurious
power spectra calculated here are less noisy because they do not include the cosmic variance. 
The formalism described here directly employs the
underlying power spectrum as opposed to Ponthieu [10] 
whose results are based on producing synthetic CMB maps and their first and second spatial derivatives with HEALPix (G{\'o}rski et al. [32]). These are combined with, e.g. the first order moments of the effective beam and hits distribution in each pixel, producing simulated `observed' maps (including their systematic effects), 
and these maps can be studied by themselves or used to produce power spectra.
Although our analytic results are given in their most general form, including the effect 
of scanning strategy, in practice $f_{1}$, $f_{2}$ and $f_{3}$ (Eq. 27) should be 
numerically calculated. This is consistently done by the approach employed by Ponthieu [10].  
Our analysis shows that the contributions from higher order corrections and coupling 
between the various effects may be of some importance in principle, but for the cases 
studied here of a perfect elliptical gaussian beam with mean beamwidth $\approx 1^{\circ}$ 
and ellipticity and pointing error on the 5\% level with a perfect scanning strategy, 
we find that the higher order corrections have a negligible effect on the angular scales of interest.
Another aspect is the potential use of the nonvanishing $C_{l}^{TB}$ and $C_{l}^{EB}$ power spectra 
as monitors of an imperfect removal of the spurious beam effects using the fact that these 
two power spectra are expected to identically vanish in the standard cosmological model. 

\ \\
{\it Acknowledgments:}\ 
We wish to acknowledge fruitful discussions with Jamie Bock and the EPIC working group.
MS would like to thank Evan Bierman, Nathan Miller, Tom Renbarger 
and John Kovac for useful discussions. 
BK gratefully acknowledges support from NSF PECASE Award AST-0548262 
and a NASA Einstein Probe mission study grant number 
785.90.00.04 ``The Experimental Probe of Inflationary Cosmology (EPIC)".
We acknowledge using CAMB to calculate the power spectra 
of the underlying sky presented in this work. 

\newpage

\appendix*
\section{Appendix A: Auxiliary Correlation Functions}

We list here the auxiliary correlation functions $A$ and $C_{\pm}$ 
\begin{eqnarray}
C_{\pm} &\equiv &\langle (E+iB)(E\pm iB)^{*}\rangle=\nonumber\\
& &\sum_{mnm'n'}\langle\tilde{h}_{+}(m,n)\tilde{h}_{\pm}^{*}(m',n')e^{i[2(m-m')+(n-n')]\phi_{l}}\rangle\nonumber\\
&\star&\left[4C_{l}^{T}(B_{-})_{m+1,n}(B_{-})_{m'\pm 1,n'}^{*}\right.\nonumber\\
&+&\left.2C_{l}^{TE}[(B_{-})_{m+1,n}\left((B_{+})_{m'n'}+(B_{+})_{m'\pm2,n'}\right)^{*}\right.\nonumber\\
&+&\left.(B_{-})_{m'\pm 1,n'}^{*}\left((B_{+})_{m,n}+(B_{+})_{m+2,n}\right)]\right.\nonumber\\
&+&\left.C_{l}^{E}[(B_{+})_{m'n'}^{*}+(B_{+})_{m'\pm 2,n'}^{*}]\right.\nonumber\\
&\times &\left.[(B_{+})_{m,n}+(B_{+})_{m+2,n}]\right.\nonumber\\
&\pm&\left.C_{l}^{B}\left((B_{+})_{m,n}-(B_{+})_{m+2,n}\right)\right.\nonumber\\
&\times&\left.\left((B_{+})_{m',n'}^{*}-(B_{+})_{m'\pm 2,n'}^{*}\right)\right.\nonumber\\
&\pm&\left.2iC_{l}^{TB}[(B_{-})_{m+1,n}\left((B_{+})_{m'\pm 2,n'}-(B_{+})_{m',n'}\right)^{*}\right.\nonumber\\
&\pm&\left.(B_{-})_{m'\pm 1,n'}^{*}\left((B_{+})_{m,n}-(B_{+})_{m+2,n}\right)]\right.\nonumber\\
&\pm&\left.iC_{l}^{EB}[((B_{+})_{m,n}+(B_{+})_{m+2,n})\right.\nonumber\\
&\times&\left.((B_{+})_{m'\pm 2,n'}^{*}-(B_{+})_{m',n'}^{*})\right.\nonumber\\
&\pm&\left.((B_{+})_{m,n}-(B_{+})_{m+2,n})\right.\nonumber\\
&\times&\left.((B_{+})_{m'\pm 2,n'}^{*}+(B_{+})_{m',n'}^{*})\right]\nonumber
\end{eqnarray}
\begin{flushright}
(A.1)
\end{flushright}

\begin{eqnarray}
A &\equiv &\langle T(E-iB)^{*}\rangle=\nonumber\\
& &\sum_{mnm'n'}\langle\tilde{f}(m,n)\tilde{h}^{*}_{-}(m',n')e^{i[2(m-m')+(n-n')]\phi_{l}}\rangle\nonumber\\
&\star&\left[2C_{l}^{T}(B_{+})_{mn}(B_{-})_{m'-1,n'}^{*}\right.\nonumber\\
&+&\left.C_{l}^{TE}[(B_{+})_{mn}\left((B_{+})_{m'n'}+(B_{+})_{m'-2,n'}\right)^{*}\right.\nonumber\\
&+&\left.(B_{-})_{m'-1,n'}^{*}\left((B_{-})_{m+1,n}+(B_{-})_{m-1,n}\right)]\right.\nonumber\\
&+&\left.\frac{1}{2}C_{l}^{E}[(B_{+})_{m'n'}^{*}+(B_{+})_{m'-2,n'}^{*}]\right.\nonumber\\
&\times&\left.[(B_{-})_{m+1,n}+(B_{-})_{m-1,n}]\right.\nonumber\\
&+&\left.\frac{1}{2}C_{l}^{B}[(B_{+})_{m',n'}^{*}-(B_{+})_{m'-2,n'}^{*}]\right.\nonumber\\
&\times&\left.[(B_{-})_{m+1,n}-(B_{-})_{m-1,n}]\right.\nonumber\\
&+&\left.iC_{l}^{TB}[(B_{+})_{mn}\left((B_{+})_{m',n'}-(B_{+})_{m'-2,n'}\right)^{*}\right.\nonumber\\
&+&\left.(B_{-})_{m'-1,n'}^{*}\left((B_{-})_{m-1,n}-(B_{-})_{m+1,n}\right)]\right.\nonumber\\
&+&\left.\frac{i}{2}C_{l}^{EB}[((B_{-})_{m+1,n}+(B_{-})_{m-1,n})\right.\nonumber\\
&\times &\left.((B_{+})_{m'n'}^{*}-(B_{+})_{m'-2,n'}^{*})\right.\nonumber\\
&-&\left.((B_{-})_{m+1,n}-(B_{-})_{m-1,n})\right.\nonumber\\
&\times &\left.((B_{+})_{m',n'}^{*}+(B_{+})_{m'-2,n'}^{*})]\right]\nonumber
\end{eqnarray}
\begin{flushright}
(A.2)
\end{flushright}
where $\star$ denotes 2-D convolution.

\end{document}